\documentclass[12pt]{article}
\usepackage[pdftex]{graphicx}
\usepackage{amsmath,amssymb}
\usepackage{hyperref} 

\begin{document}

\title{Quantum Measurements As a Control Resource}
\date{}
\author{Alexander N. Pechen\thanks{\href{mailto:pechen@mi.ras.ru}{pechen@mi.ras.ru}; \href{http://www.mathnet.ru/eng/person/17991}{www.mathnet.ru/eng/person/17991}}}
\maketitle

\vspace{-1cm}
\begin{center}
Chemical Physics Department, Weizmann Institute of Science, Rehovot 76100, Israel\\ 
Steklov Mathematical Institute, Gubkina str. 8, Moscow 119991, Russia
\end{center}

\begin{abstract}
We discuss the use of back-action of quantum measurements as a resource for
controlling quantum systems and review its application to optimal approximation
of quantum anti-Zeno effect.
\end{abstract}

\section{Introduction}
Quantum measurements are commonly used to extract information about the measured
system~\cite{Mensky2000} that can then be used to control the system via
feedback~\cite{Wiseman1993,Belavkin2009}. However, quantum measurements also
typically affect the system through the measurement's back-action that can be
directly used to manipulate the system dynamics even if the measurement results
are not recorded
\cite{Grishanin2003,VilelaMendes2003,Grishanin2004,Pechen2006,Roa2006,Erez2008}. A manifestation of such control is the quantum anti-Zeno effect, where
continuous observation of certain
time-dependent operators steers the system dynamics along a predefined
pathway~\cite{Balachandran2000,Facchi2001}. The use of the anti-Zeno effect
requires
continuous monitoring of the system that sometimes can be hard to realize. We
review the proposed in~\cite{Pechen2006} general scheme for optimal control by
discrete
quantum measurements and consider as an application optimal approximation of the
anti-Zeno effect by a finite number of measurements.

\section{Optimal Control by Quantum Measurements}
A non-selective measurement of an observable $Q$ with the spectral decomposition
$Q=\sum_iq_iP_i$, where $q_i$ is an eigenvalue and $P_i$ is the corresponding
projector, transforms the system density matrix $\rho_{\rm i}$ into ${\cal
M}_Q(\rho_{\rm i}):=\sum_i P_{i}\rho_{\rm i} P_{i}$. Measuring the
observables $Q_1,\dots, Q_N$ evolves the system state $\rho_{\rm i}$ into
\begin{equation}\label{eq1}
\rho_{\rm final}={\cal M}_{Q_N}\circ{\cal M}_{Q_{N-1}}\dots\circ {\cal
M}_{Q_1}(\rho_{\rm i})
\end{equation}
This induced by non-selective quantum measurements transformation of the system
state can be used to optimize some system related properties. One class of
practically important problems can be described by maximizing the
expectation value of a target operator $O$
\begin{equation}\label{eq2}
J_N[Q_1,\dots,Q_N]={\rm Tr}[\rho_{\rm final} O]\to\max
\end{equation}
The control goal is to find optimal observables $Q^*_1, \dots,Q^*_N$ such that
their sequential measurement maximizes $J_N$.

The back-action of quantum measurements can be supplemented by coherent control
$u(t)$ acting between the measurements to produce a unitary system dynamics
governed by the equation $\dot{\rho}(t)=-i[H_0-\mu u(t),\rho(t)]$, where $H_0$
is the free system Hamiltonian and $\mu$ is the dipole moment. We denote ${\cal
U}_i(\rho)=U_i\rho U^\dagger_i$, where $U_i$ is the unitary evolution between
the $i$-th and $(i+1)$-th measurements. Then the system density matrix after $N$
measurements will be
\[
\rho_{\rm final}={\cal U}_N\circ{\cal M}_{Q_N}\circ {\cal U}_{N-1}\circ{\cal
M}_{Q_{N-1}}\circ \dots\circ {\cal U}_1\circ{\cal M}_{Q_1}\circ {\cal U}_0
(\rho_{\rm i})
\]
The density matrix $\rho_{\rm final}$ depends on $u(t)$ and
$Q_1,\dots Q_N$ and determines the objective functional
\[
J[u(t),Q_1,\dots,Q_N]={\rm Tr}\,[\rho_{\rm final} O]\to\max
\]
where both the standard coherent field $u(t)$ and the observables
$Q_1,\dots,Q_N$ are treated as controls to be optimized~\cite{Pechen2006}.

\section{Optimal Approximation of Anti-Zeno Dynamics}
According to quantum anti-Zeno effect, under continuous measurement of the
projection
operator $E(t) = U(t)EU^\dagger (t)$, where $E$
is a
projector leaving the initial state unchanged and $U(t)$ a unitary operator such
that $U(0) = 1$, the probability of finding $E(t) = 1$ at any $t\in[0,T]$
is one~\cite{Balachandran2000}. Thus the evolution of the continuously
monitored system will follow the prescribed pathway $E(t)$. Continuous quantum
measurements
may sometimes be hard to realize and optimal approximations of the anti-Zeno
dynamics by a
finite number of measurements may be desirable. In~\cite{Pechen2006} an optimal
approximation of anti-Zeno effect by any finite number of measurements was
obtained for two-level systems. As an example, Fig.~1 shows on the Bloch sphere
the Stokes vectors for ten optimal measurements which approximate the anti-Zeno
dynamics with $E=|0\rangle\langle 0|$ and $E(T)=|1\rangle\langle 1|$.

\begin{figure}[htbp]
  \centering
  \includegraphics[width=8.3cm]{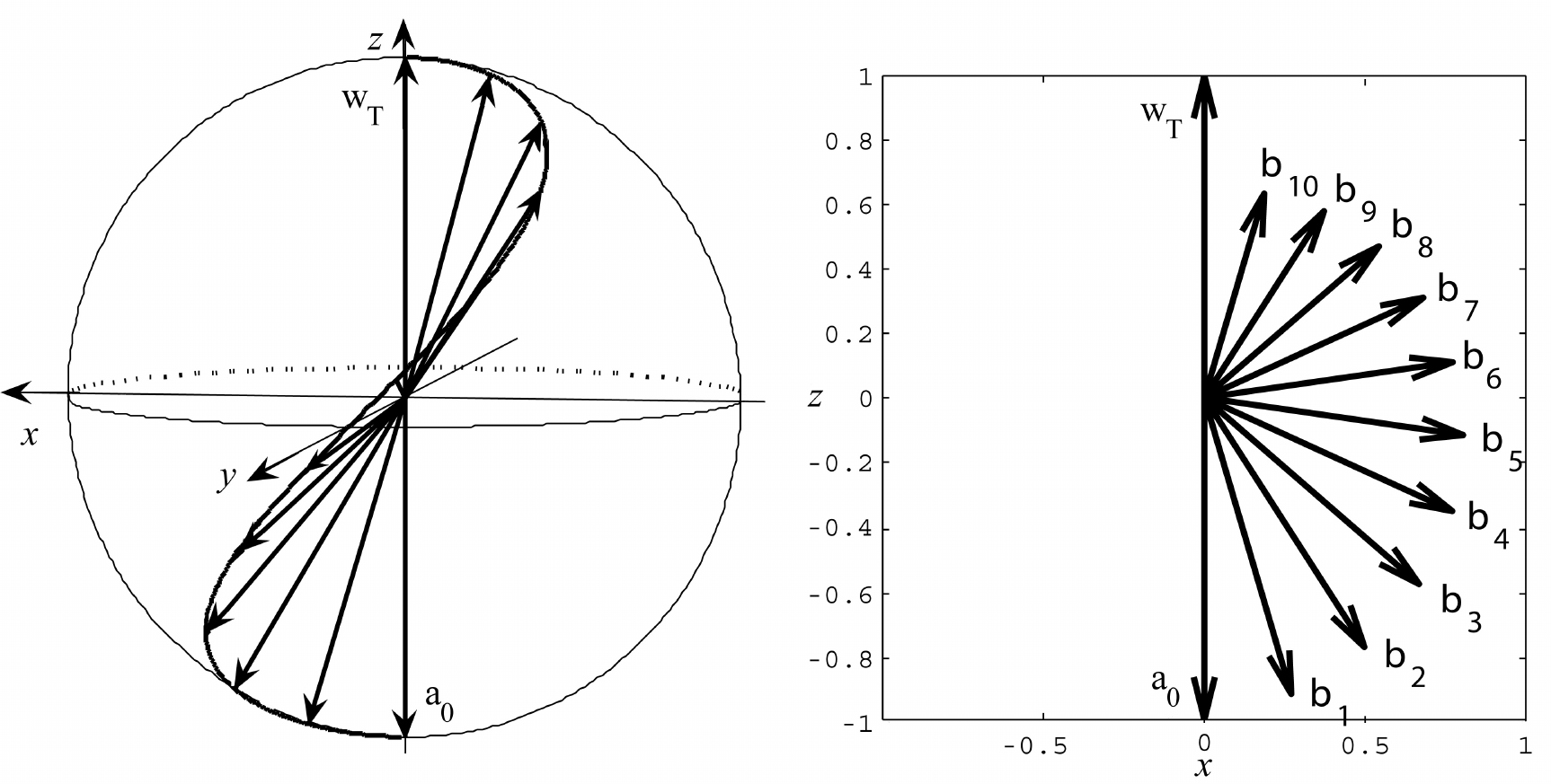}
\caption{(From Ref.~\cite{Pechen2006}. Copyright (2006) by the American Physical
Society.) Optimal approximation of the anti-Zeno dynamics in a two-level
system by ten quantum measurements. The initial system state is $\rho_{\rm
i}=\frac{1}{2}(\mathbb I+a_0\sigma)=E$ and the desired target state $\rho_{\rm
T}=\frac{1}{2}(\mathbb I+w_{\rm T}\sigma)=E(T)$, where
$\sigma=(\sigma_x,\sigma_y,\sigma_z)$. Left plot shows the Bloch sphere and ten
unit norm vectors $w_i=b_i/\|b_i\|$ which determine the observables (projectors)
$Q^*_i=\frac{1}{2}(\mathbb I+w_i\sigma)$ such that their sequential measurement
optimally approximates the anti-Zeno dynamics. State of the system after $i$-th
measurement is $\rho_i=\frac{1}{2}(\mathbb I+b_i\sigma)$, where the vectors
$b_i$ ($\|b_i\|<1$, thus each $\rho_i$ is mixed) are plotted on the right plot.}
\end{figure}

\section*{Acknowledgments}
This work was supported by a Marie Curie International Incoming Fellowship
within the 7th European Community Framework Programme.

\end{document}